# *In vivo* imaging of human cornea with high-speed and high-resolution Fourier-domain full-field optical coherence tomography

**Egidijus Auksorius,**[1,2,*] **Dawid Borycki,**[1,2] **Patrycjusz Stremplewski,**[1] **Kamil Liżewski,**[1] **Slawomir Tomczewski,**[1] **Paulina Niedźwiedziuk,**[1] **Bartosz L. Sikorski,**[3,4] **and Maciej Wojtkowski**[1]

[1]*Institute of Physical Chemistry, Polish Academy of Sciences, Kasprzaka 44/52, 01-224 Warsaw, Poland*
[2]*Equal contribution*
[3]*Department of Ophthalmology, Nicolaus Copernicus University, 9 M. Sklodowskiej-Curie St., Bydgoszcz 85-309, Poland*
[4]*Oculomedica Eye Research & Development Center, 9 Broniewskiego St, 85-391 Bydgoszcz, Poland*
*\*egidijus.auksorius@gmail.com*

**Abstract:** Corneal evaluation in ophthalmology necessitates cellular-resolution and fast imaging techniques allowing accurate diagnoses. Currently, the fastest volumetric imaging technique is Fourier-domain full-field optical coherence tomography (FD-FF-OCT) that uses a fast camera and a rapidly tunable laser source. Here, we demonstrate high-resolution, high-speed, non-contact corneal volumetric imaging *in vivo* with FD-FF-OCT that can acquire a single 3D volume with a voxel rate of 7.8 GHz. The spatial coherence of the laser source was suppressed to prevent it from focusing to a spot on the retina, and therefore, exceeding the maximum permissible exposure (MPE). Inherently volumetric nature of FD-FF-OCT data enabled flattening of curved corneal layers. Acquired FD-FF-OCT images revealed corneal cellular structures, such as epithelium, stroma and endothelium, as well as subbasal and mid-stromal nerves.

## 1. Introduction

Corneal imaging is crucial in the diagnosis and treatment of eye diseases. Imaging of cornea, however, is challenging due to its relatively high transparency in the visible and near-infrared spectral regions. Slit-lamp biomicroscopy and specular microscopy are routinely employed to image epithelium and endothelium layers of the cornea, but they cannot visualize its inner structures, such as stroma, due to the lack of optical sectioning. To address that, confocal microscopy has been introduced [1] to discriminate different corneal layers by the help of confocal pinhole that provides optical sectioning. However, it is a relatively slow technique since volumetric (3D) imaging requires point-by-point scanning of the cornea in all three directions (*xyz*), which makes the method vulnerable to imaging artifacts due to eye motions. Nipkow disc [2] or line scanning [3] confocal microscopy can be used to achieve higher speeds. However, to efficiently resolve different corneal layers axially, high numerical aperture (NA) objectives have to be used to create a limited depth-of-field (DOF), which, however, typically requires immersion liquid. Although the immersion also helps to reduce specular reflections from the cornea, it does, however, necessitates ocular anesthesia, making the confocal microscopy a contact method and, overall, less attractive for routine medical examinations.

Optical Coherence Tomography (OCT) is a well-established tool in corneal imaging [4, 5]. It has the advantage of axial resolution being decoupled from the lateral that enables achieving relatively high axial resolution in a non-contact manner with a low NA objective lens if a spectrally broadband light source is used. Furthermore, volumetric data can be recorded faster with Fourier-domain (FD) OCT than with confocal microscopy since axial information (A-scans) in FD-OCT can be acquired almost instantaneously – without physically scanning along the optical axis. OCT also has the sensitivity advantage over the confocal microscopy because it combines spatial gating – as implemented by the confocal pinhole – with the temporal gating [6]. The combination allows deeper imaging through thick layers of scattering media due to a more efficient rejection mechanism of multiply scattered light than in confocal microscopy. OCT has been continuously improving in terms of speed [4, 7] and axial resolution. Use of spectrally broad and spatially coherent light sources, such as supercontinuum lasers, was critical in bringing the axial resolution down to 1 μm [8, 9] for corneal imaging *in vivo*. Such high axial resolution resolved endothelial cells axially (in B-scans) but failed to do it laterally – in the *en face* view – due to the poor lateral resolution that fell around 20 μm. Higher NA objective lenses were employed to get lateral resolution closer to that of axial [10]. Finally, *en face* imaging of endothelium cells *in vivo* was demonstrated in animals [11, 12] and humans [13]. Nevertheless, to achieve such high spatial resolution, it requires faster data acquisition to avoid significant image blur due to eye motion.

Potentially faster alternatives to the point scanning OCT are line-field OCT [14] and full-field OCT (FF-OCT) [15, 16], where a sample is illuminated with a line or wide-field illumination that allows parallelized signal detection with a 1D or 2D detector, respectively. Conventional FF-OCT can perform fast *en face* imaging [15, 16] and is based on a camera and spatially incoherent illumination source, such as LED. We call this particular FF-OCT technique – TD-FF-OCT due to its time-domain (TD) signal acquisition, to distinguish it from a Fourier-domain variant – FD-FF-OCT, described below. However, to generate 3D volumes, a sample (or reference mirror) in TD-FF-OCT needs to be translated axially, making it a slow volumetric imaging technique. On the other hand, high-NA objectives can be used with the axial sample scanning to achieve better than 1 μm isotropic resolution [15, 16]. Meanwhile, in most of other OCT techniques the lateral resolution has to be balanced against the DOF, which effectively defines the axial imaging range [6, 17]. The use of TD-FF-OCT for *ex vivo* eye imaging [18-21] and *in vivo* imaging of anesthetized rat's eye [22] has been demonstrated a while ago. Only recently, however, it has been successfully employed for the *in vivo* imaging of human cornea [23], as well as retina [24, 25]. The eye imaging applications were spearheaded by the development of fast and high full-well-capacity (FWC) camera that, for example, demonstrated sensitivity of >100 dB (in 1 s acquisition) without pixel binning [26]. Despite the fast *en face* (2D) image acquisition and camera development efforts, TD-FF-OCT remains relatively slow volumetric (3D) imaging technique, which limits its use in high-resolution eye imaging. The concept of Fourier-domain FF-OCT (FD-FF-OCT) has been introduced in the past that can acquire volumetric images faster [27], but only recently it reached its potential when high-speed cameras with frame rates of tens of kilohertz were employed [28-30]. FD-FF-OCT, otherwise known as Full-Field Swept-Source OCT (FF-SS-OCT), also utilizes a tunable laser source, usually referred to as a swept laser source. Such a system can record interferometric images as a function of wavelength (frequency, $\omega$) – forming an interference spectrum for each image pixel. Fourier-transform ($\omega \rightarrow z$) of such spectrum in each pixel gives the axial profile, $z$ of a sample – an A-scan. Fourier-domain signal detection and its parallelization by such fast cameras enable ~10 GHz voxel rate [28], making FD-FF-OCT the fastest volumetric OCT technique with relatively high sensitivity. Even though the same volumetric imaging speed is possible with TD-FF-OCT [31], the achieved sensitivity can be orders of magnitude lower [32-34]. Also, TD-FF-OCT involves moving a sample or parts of the system for 3D data acquisition, whereas FD-FF-OCT has no moving parts. Besides, phase information is available in Fourier-domain methods (scanning or full-field) that can be used to correct not only for the chromatic dispersion between the arms but also for the axial motion, which appears as a chirp in recorded interference spectrum and is inevitable in imaging *in vivo*. The phase in FD-FF-OCT is stable across the whole field of view, and therefore, is not corrupted by the sample motion, like it is typically between A-scans in scanning OCT. The phase stability in FD-FF-OCT permits computational aberration compensation that can enhance image quality [28]. Another important full-field imaging benefit is that considerably higher power exposure is allowed on eye in both time- and Fourier-domain systems compared to the scanning OCT due to the power distributed across larger area [35]. FD-FF-OCT so far has been exclusively used for the retinal imaging [28-30, 36-38], except one skin imaging demonstration [39]. Cornea arguably can be a good target for FD-FF-OCT because of low scattering that does not necessitate confocal detection. However, imaging with spatially coherent FD-FF-OCT poses a safety risk to the retina. Namely, a collimated laser beam used to illuminate the cornea is focused to a spot on the retina by the cornea and the eye lens, which can easily exceed the maximum permissible exposure (MPE) limit. Even though, other illumination schemes are possible (even in the transmission mode [40]), nevertheless, sending the collimated beam on the cornea seems to be the most optimal since the cornea backscatters light predominantly in the forward/backward directions. To this end, we use here a spatially incoherent swept laser source, which we prepare by destroying the spatial coherence of the laser with a fast-deformable membrane (DM). Such incoherent illumination effectively broadens the laser spot on the retina to the acceptable MPE intensity levels. We have previously used DM to remove crosstalk in FD-FF-OCT images of skin [39] and retina [41]. In corneal imaging, however, crosstalk is less of an issue due to the low corneal scattering, and therefore, the DM is primarily used to circumvent the laser safety problem. We have also performed digital aberration correction (DAC) on the acquired 3D data volumes, as described in Ref. [42], which helped to correct for defocus. We thus were able to show that volumes, acquired in a fraction of a second and averaged to increase the signal-to-noise ratio (SNR), allowed capturing a significant portion of the cornea. The images showed various corneal layers with good delineation, including that of the endothelium cell layer and nerve plexuses.

## 2. Material and Methods

### 2.1 FD-FF-OCT system

The main components of the imaging system, shown in Fig. 1, were a fast-tunable laser source, a deformable membrane (DM), a Linnik interferometer and a fast camera. A tunable laser source (Broadsweeper BS-840-2-HP, Superlum) – capable of delivering 25 mW of average output power – could be tuned from 800 nm to 878 nm at a sweeping rate of up to 100 000 nm/s. The light from the source was delivered to a setup by a single-mode fiber. It was collimated by an objective lens and a *4-f* telescope (not shown) to a diameter of 1.65 mm as measured at the full-width-half-maximum (FWHM). The beam was then reflected off by the DM (Dyoptyka), where a sequence of phase patterns – largely uncorrelated to each other – was imprinted on its wavefront in ultra-fast succession. DM was manufactured from a highly reflective thin membrane that could form quickly changing standing wave patterns. Each of the patterns was effectively acting as a dynamic random diffraction grating diffracting the impinging beam into a range of angles. When DM was switched off, lens, *L1* was focusing the beam to an almost diffraction-limited spot of 22 μm at a focal plane *P'* that was conjugate to the pupil plane, *P* of the objective lens ($f = 1.8\ cm$), *L4*, as shown in Fig. 1. Upon switching the DM 'on', the spot was be broadened to the FWHM size of 630 μm, as shown in the inset of Fig. 1. The *P'* plane was further relayed by the lenses *L2* and *L3* – arranged in the *4-f* configuration – on the pupil plane, *P*. The relay lenses provided ×2 demagnified image of the spot on the pupil plane. They were necessary to produce the spot size in the pupil of the objective that otherwise would have required $f = 2.5\ cm$ lens, which could not be put into the *4-f* configuration with the objective lens due to the mechanical constraints. It would have also required DM to be tilted by a large angle. The 50/50 beamsplitter split the beam into the reference and sample arms. The reference arm comprised an objective lens, *L4* (UPlanFL N, Olympus, ×10, NA = 0.3, FN = 26.5), a mirror and a tilted neutral density (*ND*) filter that attenuated the reference beam to ~10 % in the double-pass configuration.

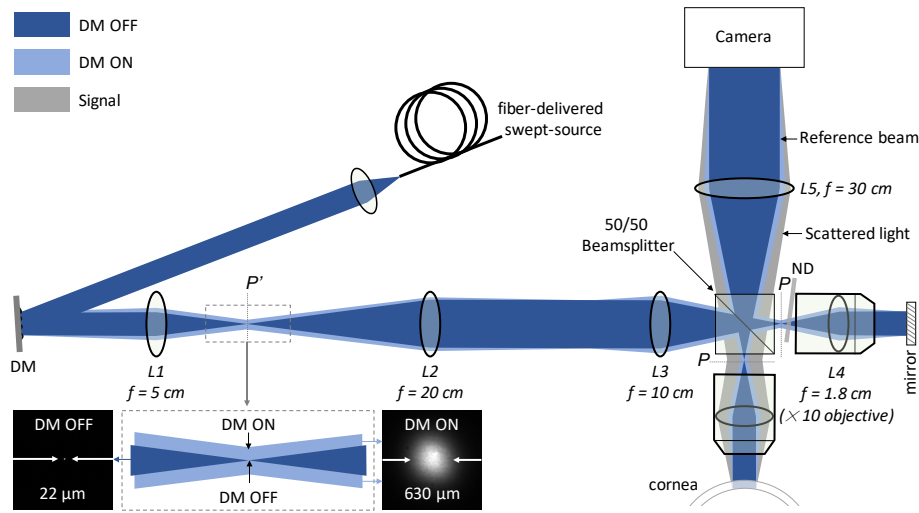

Fig. 1. Fourier-domain full-field OCT system for *in vivo* corneal imaging of the human eye. DM – deformable membrane; ND – neutral density filter; P' – plane conjugate to the pupil plane, P of the objective lens, L4; Dark blue shows the spatially coherent beam (when DM is OFF) and light blue – the spatially incoherent (when DM is ON). Gray beam indicates signal coming from the cornea that is backscatter at a range of angles. The beam is focused to a spot of 22 μm in the P' plane in the coherent case (when DM is OFF), which is broadened to 630 μm in the incoherent case (when DM is ON), as shown zoomed-in in the inset.

The sample arm contained the same type objective lens, *L4*. The objective lenses collimated beams to the FWHM diameter of 1.2 mm when the DM was switched off. Switching DM 'on' enlarged the spot size in the pupil plane from 11 μm to 315 μm, which produced a range of collimated beams with the same diameter of 1.2 mm traveling at different angles with respect to the optical axis. A collection of such beams resulted in the illumination with a convergence angle of 0.5° going onto the cornea and reference mirror, as can be seen in Fig. 1. Field-of-view (FOV) did not change. Backscattered light from the cornea and reflected light from the reference arm was recombined by the same beamsplitter and imaged on the camera by the tube lens, *L5* (*f = 30 cm*), resulting in the magnification, *M* of 30 cm/1.8 cm = ×16.7. An interference image was detected by a fast camera (Fastcam SA-Z, Photron) that was able to acquire images with 1024×1024 pixels at 20,000 frames-per-second (fps), or faster if images had fewer pixels. It featured 20 μm pixels that sampled cornea every 1.2 μm at this particular magnification (of ×16.7). DM, in turn, was effectively demagnified 1.39 times on the cornea.

The system was designed such that the reflections in the system were minimized and prevented from reaching the camera. This was necessary despite all the optical elements being anti-reflection coated to reduce incoherent light. To this end, the DM was mounted on the tip-and-tilt stage that allowed shifting the beam laterally in the pupil plane, *P* of the objective lenses, *L4*. This enabled to direct the specular reflections from the objective lenses away from being detected by the camera due to the concave shape of the first lens in the objective on its pupil side. The beamsplitter was rotated in-plane by a small angle (not shown in Fig. 1) to avoid specular reflection from the beamsplitter going on to the camera, which also required the illumination arm to be built at an angle.

*2.2 Illumination on the retina*

To maximize the backscattered light coming from the cornea, and thus, the OCT signal, it was illuminated with a collimated laser beam. However, such a spatially coherent beam was focused to a spot on the retina by the cornea and the eye lens exceeding the maximum permissible exposure. Therefore, when using a collimated laser with a few milliwatts of power, it can only be below the safety limit for the cornea, but not for the retina.

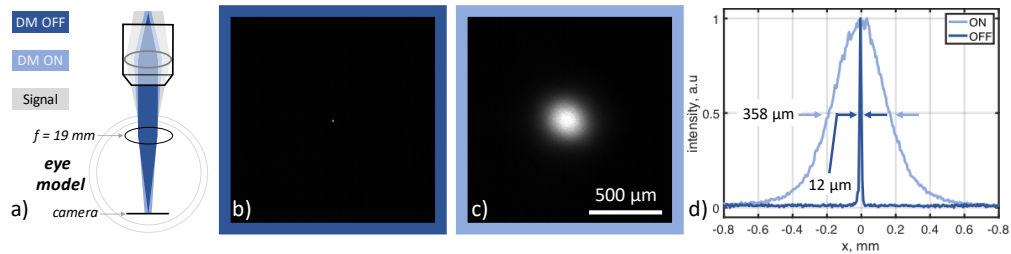

Fig. 2. Increasing the size of the laser beam on the retina to circumvent safety problem. The size of the laser spot on the retina in the eye model with (c) and without (b) dynamic phase modulation. (a) A sample arm – part of the setup shown in Fig. 1 – and an eye model. It demonstrates how a laser beam is focused on the retina in the coherent (DM OFF) and incoherent (DM ON) cases. For the eye model, an achromatic doubled ($f = 19\ mm$) was used. An image of the focused spot was recorded by a camera placed at the focal plane of the lens corresponding to the retina plane in the eye model when DM was switched OFF (b) and ON (c). The size of the focused spot was ~ 12 μm in (b) and was broadened to 358 μm in (c), as can be seen from the line profiles, shown in (d). The illumination radius increase is ~30 times, and the area – 890 times – greatly reducing the laser intensity on the retina.

Here we circumvented the safety problem by using DM that made the laser beam spatially incoherent due to the phase randomization, which effectively broadened the focused spot size ~30 times (Fig. 1). To illustrate this effect on the retina, an eye model was built and inserted in Fig. 1 system. The eye model, shown in Fig. 2(a), consisted of a $f = 1.9\ cm$ lens and a camera positioned at the focal length of the lens. A spatially coherent laser produced a 12 μm spot on the camera, as shown in Fig. 2(b), whereas activating DM broadened the spot ~30 times – to ~360 μm, as shown in Fig. 2(c). Thus, the intensity of the focus dropped by almost three orders of magnitude reducing the retinal irradiance from 1263 W/cm$^2$ to 1.54 W/cm$^2$. In contrast to our previous work, where phase randomization was used to reduce crosstalk in retinal images, here it was mainly exploited to circumvent the laser safety issues.

*2.3 Corneal imaging procedure in vivo*

For the corneal imaging *in vivo*, we acquired images with 512 × 512 pixels at the speed of 60,000 fps. We recorded 512 images while tuning the laser with a speed of 8700 nm/s that resulted in the acquisition of 116 volumes per second (8.6 ms per volume). We typically acquired 30 such volumes resulting in the total acquisition time of 260 ms. For the endothelium cell imaging, we acquired 1024 images per volume, which resulted in 17.2 ms volume acquisition time. However, 512 image acquisition could equally be used instead of 1024 to shorten the time. The volunteer's eye was imaged with the system, shown in Fig. 1, using only a standard chin-and-head rest to stabilize the volunteer's head. The chin-and-head rest was mounted on the 3-D translation stage that allowed adjusting the position of the head axially, as well as laterally. DM was always switched on so that the beam would not be focused to a spot on the retina exceeding MPE, as explained in section 2.2. For imaging, the position was adjusted until the OCT signal of the cornea was detected in the preview mode running at a few fps. This real-time cross-sectional visualization helped in cornea alignment. The power sent on the eye was <5 mW, which was lower than MPE for the spatially extended illumination. The images were acquired in a non-contact fashion with a distance of 1 cm between the cornea and the microscope objective. Specular microscopy – a gold standard in corneal imaging – was also used to acquire images of endothelium with a specular microscope (XV NSPC, Konan, South Korea) for comparison purposes. The imaging was conducted in accordance with the tenets of the Declaration of Helsinki. Written informed consent was obtained from all subjects prior to OCT imaging and after explanation of all possible consequences of the examination. The study protocol was approved by the ethics committee of the Collegium Medicum of Nicolaus Copernicus University, Bydgoszcz, Poland.

## 3. Image processing

The processing pipeline proceeds similarly to the one used for retinal imaging [41]. However, we are now using generalized spatial filtering, automatic dispersion compensation, digital aberration correction, and automatic corneal curvature correction. We also perform automatic endothelial cell segmentation based on artificial neural networks, as explained below.

*3.1 Fourier transformation*

Each spectral A-scan from the input volume $(x, y, \omega)$ is first bandpass filtered to remove the DC level. The resulting signal is resampled to ensure the fringe linearity, and zero-padded with either 1536 or 3072 zeros (for 512 or 1024 image acquisition modes, respectively). Finally, the Fourier transformation ($\omega \rightarrow z$) yielded the volumetric complex (amplitude and phase) representations of the sample $(x, y, z)$.

*3.2 Spatial filtering*

The resulting FD-FF-OCT volumes were spatially filtered to suppress the fixed pattern noise. For the retinal imaging [41], we used a simplified version of this procedure, in which low spatial frequencies were masked with a circle of the fixed radius. Here, as sketched in Fig. 3, the spatial filtering proceeds with 2D Fourier transformation for each layer to achieve the spatial spectrum.

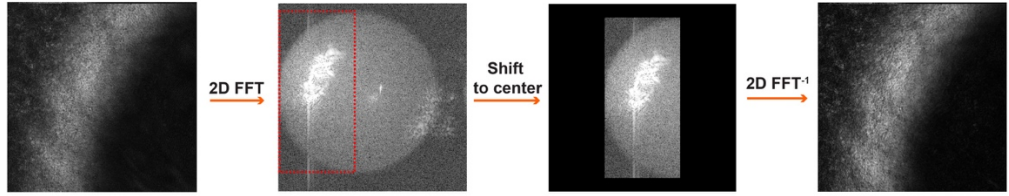

Fig. 3. Spatial filtering of FD-FF-OCT volumes. The input layer is first 2D Fourier transformed. The selected region of the spatial frequencies is shifted to the center, and then inverse-Fourier transformed to suppress the fixed pattern noise.

It is followed by a manual choice of the rectangular region that contains signal-carrying spatial frequencies. A filtered portion of the reciprocal space is shifted to the center and inverse Fourier-transformed to obtain the corrected field. Though this procedure slightly reduces the spatial resolution, we found it to be particularly important for enhancing the contrast of the endothelium cells.

*3.3 Chromatic dispersion and axial motion correction*

Spatial filtering is followed by the chromatic dispersion mismatch correction, as well as correction of eye axial motion that happens during the laser sweep in each volume [37]. We perform the correction following the method described in Ref. [43]. Namely, the $xz$ or $yz$ complex planes extracted from the center of the spatially filtered volume are Fourier-transformed, and then multiplied by the phase factor, $e^{i[a_2(\omega-\omega_0)^2+a_3(\omega-\omega_0)^2]}$ where $a_2, a_3$ are adjustable coefficients. The phase-corrected data are then inverse Fourier-transformed. This process is continued until the magnitude of the corrected field became sharp. To quantify this sharpness we employed kurtosis, and to improve our estimations we typically evaluate the sharpness metric on ten averaged neighboring B-scans. The estimated coefficients $a_2, a_3$ are then utilized to correct all other $xz$ or $yz$ planes.

*3.4 Digital aberration correction*

We perform the digital aberration correction (DAC) [42, 44] that proceeds, as sketched in Fig. 4. The complex data (amplitude and phase) representing the layer at a depth $z_l$, $U(x, y, z = z_l)$ of the sample is 2D Fourier transformed to achieve the spatial spectrum, $\widetilde{U}(k_x, k_y, z = z_l)$.

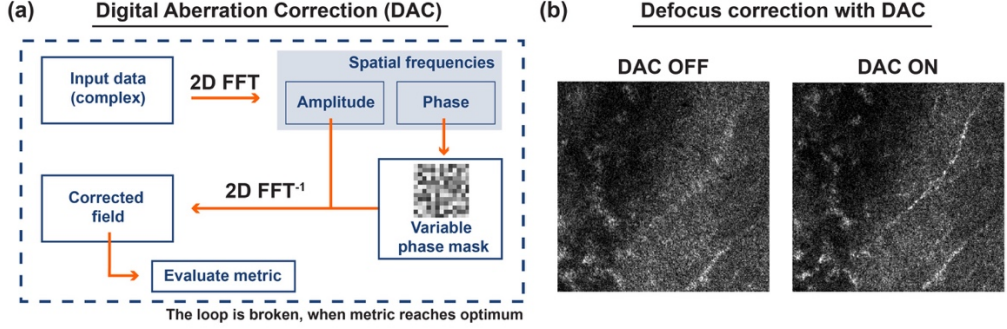

Fig. 4. Digital aberration correction (DAC). (a) The processing pipeline, in which the complex input is Fourier transformed to achieve the spatial spectrum. The latter is multiplied by the variable phase mask, and such modified spatial spectrum is inverse Fourier-transformed to obtain the corrected field $U_c(x, y)$. We then evaluate the image quality metric on $|U_c(x, y)|^2$. This process is continued until we find the metric optimum. (b) Sample results in which DAC improves image sharpness and enables us to see otherwise invisible sample features.

The latter is multiplied by the variable phase mask, $M(k_x, k_y)$ and the resulting modified spatial spectrum $\tilde{U}(k_x, k_y, z)M(k_x, k_y)$ is inverse Fourier-transformed to achieve the corrected field $U_c(x, y, z = z_l)$. We then evaluate the image quality metric on $|U_c(x, y, z = z_l)|^2$. Here, for the metric, we use the kurtosis. The phase mask is $M(k_x, k_y, z_l) = \exp[i\alpha(z_l)Z_2^0(k_x, k_y)]$, where $\alpha(z_l)$ is the depth-dependent adjustable parameter, and $Z_2^0(k_x, k_y)$ is the Zernike polynomial corresponding to the defocus (OSA/ANSI index of 4). As the sample-induced aberrations can vary laterally and axially, we partition each input volume into four sub-volumes. The sub-volumes overlap sideways with approximately 70-80 pixels. Then, we manually select the axial range, the same for each sub-volume, that includes 70-100 layers. Subsequently, for each segment in this sub-axial range, we perform DAC and store the optimum defocus value $\alpha(z_l)$ in the vector. Last, we achieve a linear fit to $\alpha(z_l) = a \times z_l + b$, and the resulting fitting parameters, $a, b$ are employed to extrapolate the depth-dependent defocus for all layers in the sub-volume, as shown in Fig. 5.

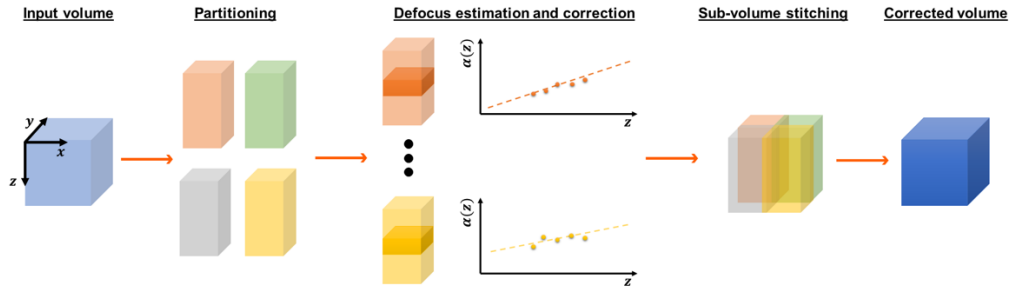

Fig. 5. Defocus correction for the entire volume. The input volume (light blue cube) is partitioned into four sub-volumes (orange, green, gray, and yellow cuboids). For each sub-volume, the depth-dependent defocus parameter $\alpha(z)$ is estimated using the manually selected axial range (shaded regions in the third column). Then each sub-volume is corrected and stitched to obtain the final defocus adjusted volume (dark blue cube).

We repeat this procedure for all the other sub-volumes, after which they are all stitched together. All volumes in the data set are processed as described above. The processed volumes are then registered (spatially aligned) and integrated.

*3.5 Cornea curvature correction*

Finally, we correct volumes for the cornea curvature by flattening them up, as sketched in Fig. 6. To this end, we first estimate the two-dimensional surface that approximates the cornea curvature. To do so, we take the central $xz$ and $yz$ B-scans, and then automatically find the most reflecting layers by locating the maximum pixel value within each A-scan. The median absolute deviations reject eventual outliers, and the resulting data is fit by the second-order polynomial. The resulting polynomials obtained for $xz$ $[p^{xz}(x,y)]$ and $yz$ B-scans $[p^{yz}(x,y)]$ are combined to obtain the surface $S(x,y)$ as follows: $S(x,y) = p^{xz}(x,y)' + [p^{yz}(x,y)]$. Symbol $'$ denotes transposition. Given the $S(x,y)$, we shift each A-scan accordingly. So, as shown in Fig. 6(b) cornea layers can be separated and visualized over the available field of view.

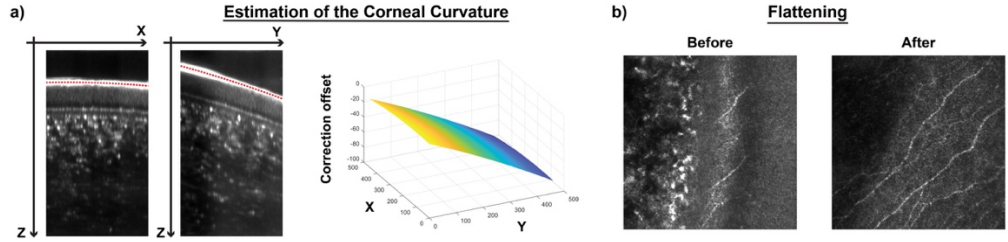

Fig. 6. Flattening volume. (a) The central $xz$ and $yz$ B-scans are used to estimate the corneal curvature by automatically finding the most reflecting layer (red dashed lines). (b) The *en face* image before flattening includes multiple cornea layers. Thus, disabling visualization of the corneal layers over the available field of view.

*3.6 Endothelial cell segmentation with neural networks*

To enable automatic image segmentation of endothelial cells, we have employed a U-net neural network that was trained by feeding it with FD-FF-OCT and specular microscopy images of endothelial cells. First, FD-FF-OCT and specular microscopy images were preprocessed by enhancing contrast with a contrast limited adaptive histogram equalization, which was followed by compensating for the non-uniform intensity distribution with flatfield correction. To this end, we divided input images by a low-pass background, which was obtained by applying a Gaussian blur on input images. Next, we normalized images using a method developed for endothelial images [41]. Finally, we smoothed images using a Gaussian blur.

We applied the U-Net neural network adopted for the endothelium cells [42]. We introduced minor modifications to the original concept but the architecture of the neural network followed a standard fully convolutional structure of the U-net, with a contraction and an expansion path, acting as encoder and decoder blocks. The output of the neural network comprises two pixel-wise probability maps for the cell edges and cell bodies, respectively. Finally, we prepared 12,000 image patches with a batch of 32 in size for the training.

Segmentation was retrieved indirectly from the network output. The trained network was then applied on a test image in a sliding window ($32 \times 32$ pixels) that was moved across consecutive patches, as explained in Ref. [42]. The final edge probability map was created as a result of averaging a network response over the overlapped patches.

After the segmentation, we use a common metrics to determine the morphometric parameters of segmented cells: the cell density (CD), the coefficient of variation (CV), and the hexagonality (HEX). CD is the total number of cells divided by the sum of individual cell areas: $CD = \frac{n_{cells}}{\sum_{i=1}^{n} S_i}$, CV is the fractional standard deviation of all cell areas: $CV = 100\% \frac{1}{\bar{S}} \sqrt{\frac{1}{n}\sum_{i=1}^{n}(S_i - \bar{S})^2}$ and HEX is the fraction of the hexagonal cells (the cells being neighbors to six other cells): $HEX = 100\% \frac{n_{hex}}{n_{cells}}$.

## 4. Results

### 4.1 System characterization

The performance of the FD-FF-OCT system for corneal imaging is summarized in Fig. 7, where spatial resolution is measured together with the Spatial Coherence Gating (SCG) imposed by the DM. Fig. 7(a) shows an image of the USAF resolution target that exhibits certain coherent artifacts appearing due to the multiple reflections of the coherent laser beam in the system. The artifacts largely disappear, as can be seen in Fig. 7(b), when the laser is made spatially incoherent by switching the DM 'on'. The FOV of the system was 615 μm × 615 μm when the camera was acquiring 512 × 512 images. A zoomed-in image, shown in Fig. 7(c), demonstrates that bars spaced 2.19 μm apart in element #6 of the group #7 (228.1 lpm) can be still resolved. Fig. 7(d) shows that a line profile over the bars dips below 50% between the bars. However, due to slight undersampling, the real lateral resolution is estimated to be 2.4 μm since the camera samples the cornea with the sampling rate of 1.2 μm/pixel, and the Nyquist criterion requires two-pixel sampling per resolution limit. Fig. 7(e) displays temporal and spatial coherence gating curves as a function of sample defocus. For these measurements, a mirror was used as a sample. The temporal coherence function was derived by Fourier-transforming interference signal recorded as a function of wavelength. The FWHM of the temporal coherence function, which corresponds to the axial resolution, was estimated to be 5.6 μm (4.2 μm in tissue) at the 100 μm from the zero temporal delay position. The peak intensity of the temporal coherence function varied as a function of the defocus when DM was switched on, as shown in Fig. 7(e) by the red curve. The intensity envelope shows how quickly light that is backscattered from a sample decorrelates axially with the defocus, which we call SCG function [41, 45]. The FWHM of the curve was estimated to be 400 μm from Fig. 7(e). It means that the light scattered outside this axial region will be randomized and not able to generate an interference pattern. It is desirable that the whole thickness of the cornea is within that value so that its entire axial range can be efficiently imaged.

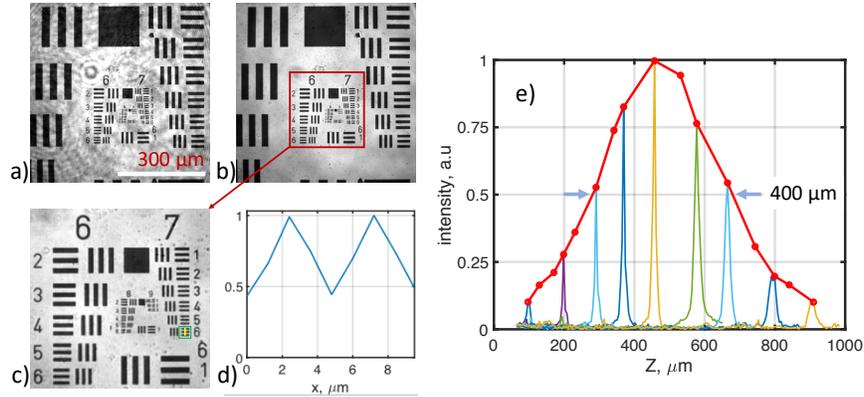

Fig. 7. System characterization. FF-OCT images of USAF target as imaged with (a) a coherent laser beam (DM OFF) and (b) incoherent beam (DM ON). Zoomed-in FF-OCT image (c) shows that element # 6 of group #7 (green square) can be resolved, as confirmed by the line profile (d) that was taken along the yellow line in (c). (e) temporal and spatial coherence gating curves representing the OCT signal as a function of the sample's defocus.

To estimate DM scattering properties, an intensity image was recorded by a separate camera placed in the focal position (*P'* in Fig. 1) of the *L1* lens. The lens focused the beam to a 22 μm (FWHM) spot when the DM was switched 'off' ('DM OFF' in the inset of Fig. 1) since it acted as a simple mirror. The spot was broadened to 630 μm when DM was activated ('DM ON' in the inset of Fig. 1). The FWHM average angle was then estimated using simple trigonometry: $arc\tan\left(\frac{0.315\,mm}{50\,mm}\right) = 0.36°$, when DM was operating at the frequency of 431.3 kHz.

## 4.2 Anterior corneal imaging in vivo

An eye of a healthy 44-years-old volunteer was imaged with the system, shown in Fig. 1, and with the procedure explained in section 2.3. Fig. 8 shows FD-FF-OCT images acquired from the anterior part of the cornea that was derived by averaging 30 volumes acquired in 260 ms with 512 multispectral images recorded per volume. Data processing was applied to the recorded data as explained in section 3. Aberration correction involved only defocus correction. An averaged B-scan, shown on the left of Fig. 8, reveals different corneal layers where tear film, epithelium and stroma layers can be identified.

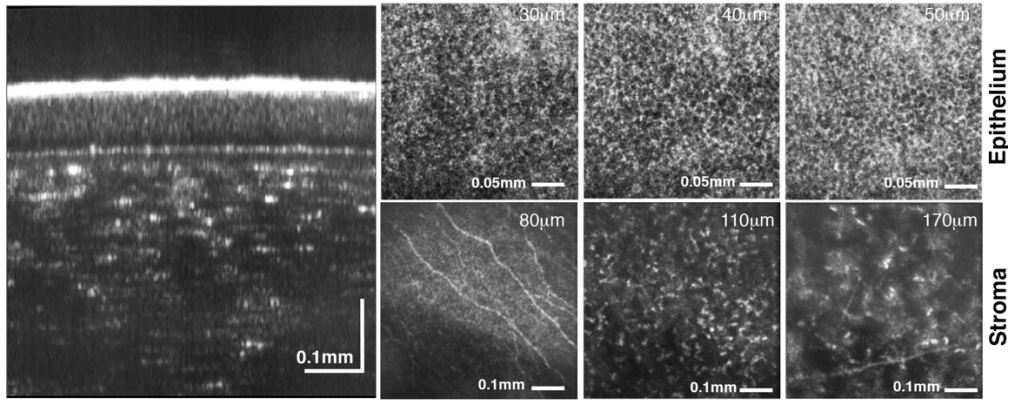

Fig. 8. Axial (left) and *en face* (right) images of human cornea acquired *in vivo*. B-scan (axial) image of cornea shown on the left displays different corneal layers. *En face* images on the right shows corneal epithelium (at 30 µm, 40 µm and 50 µm), subbasal nerves (at 80 µm), stromal nerves (at 170 µm) and anterior stroma (at 110 µm). Numbers in micrometers correspond to the depth position with respect to the corneal apex. Fly-through movies of B-scans and C-scans (*en face* images) are shown in Visualization 1 and Visualization 2, respectively

*En face* FD-FF-OCT projections of different corneal layers are also shown on the right of Fig. 8, as derived from averaged volumes. Individual epithelium cells can be seen in Fig. 8 between 30 µm and 50 µm below the corneal surface. Subbasal nerves can be clearly identified at the depth of 80 µm and stromal nerves at 170 µm. Anterior and middle stroma is also visualized with some clear keratocyte cells. Figure 9 shows snapshots at three different angles of the averaged 3D volume that was created by rotating it with FluoRender software (http://www.sci.utah.edu/software/fluorender.html) and shown in Visualization 3.

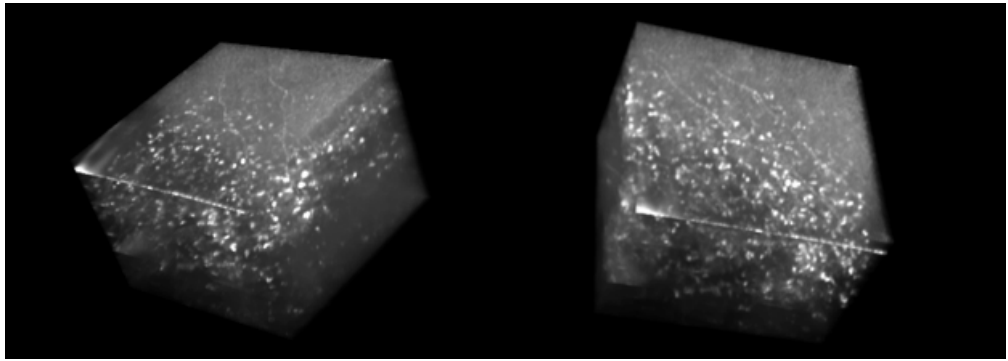

Fig. 9. 3D cutaway view (Visualization 3) of the human cornea acquired *in vivo* with FD-FF-OCT; FluoRender.

*4.3 Corneal endothelium imaging in vivo*

For the endothelium imaging, the translation stage was adjusted axially, so that the focal plane was set on the endothelial layer. Fig. 10 and Fig. 11 show non-averaged and averaged *en face* images of endothelium, respectively, as acquired with FD-FF-OCT from a healthy 43-years-old volunteer. Fig. 10 also compares the FD-FF-OCT image to a specular microscopy image of the same eye. For each volume 1024 images were acquired in 17.2 ms. FD-FF-OCT *en face* images in Fig. 10 are shown as derived from a single volume and in Fig. 10 – fifteen averaged volumes (acquired in the total time of 240 ms).

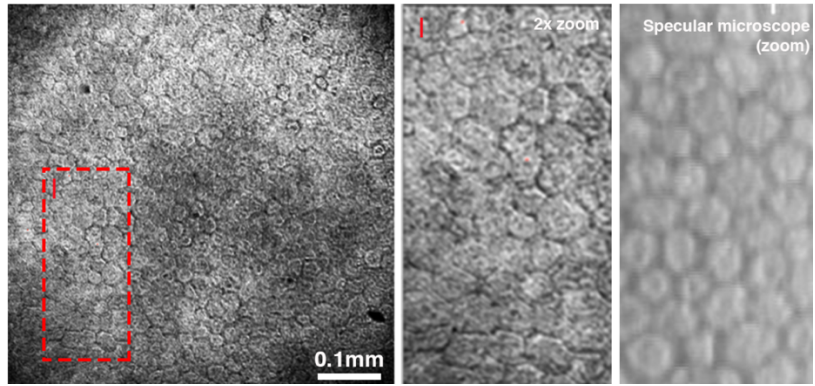

Fig. 10. Endothelium cell images derived from a single FD-FF-OCT volume acquired in 17.2 ms (left). A zoomed-in FD-FF-OCT image (middle) is compared with the specular microscopy image (right) that was recorded from the same eye and similar location.

A zoomed-in averaged *en face* FF-OCT image, shown in Fig. 11, reveal a slightly irregular shape of the endothelium cells. This could be explained by the fact that the typical hexagonal shape of the endothelium cells is only seen at one imaging plane, whereas below that plane, the shape is less regular [46]. Since the axial resolution of FD-FF-OCT is 5.6 µm, and we average 15 volumes that might be not perfectly axially aligned even after the correction, the images shown in Fig. 11 is a result of integrating regular and irregular structure in one image. Visualization 4 illustrates axial scanning of the endothelium cells as implemented computationally by refocusing. Endothelial nuclei can also be discerned in Fig. 11 zoomed-in images as dark spots.

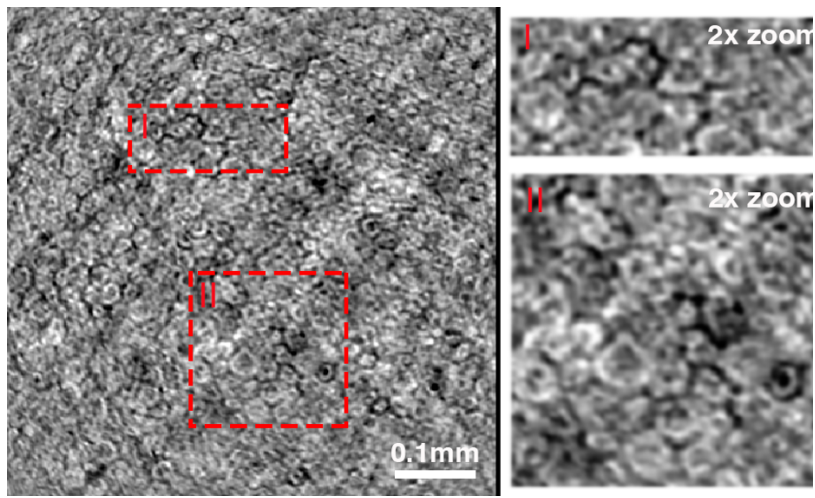

Fig. 11. Averaged endothelial cell images derived by averaging 15 volumes acquired in the total time of 240 ms. Zoomed-in images show nuclei in cells as dark spots.

*4.4 Segmentation and endothelium cell density analysis*

Segmentation of FD-FF-OCT and specular microscopy images were also performed and compared, as shown in Fig. 12. FD-FF-OCT images were segmented with a trained neural network, as explained in section 3.6. Specular microscopy images were segmented with software provided by Konan. To train the neural network, two FD-FF-OCT and four specular microscopy images were employed. For each of the images, the ground truth reference values were obtained with a manual segmentation performed by experts using a vector graphics software (Inkscape 0.92, https://inkscape.org/). The datasets consisting of FD-FF-OCT and specular microscopy images with their corresponding ground-truth segmented images were used to train the CNN network. The cell statistics derived in terms of cell density and morphometry is shown in Fig. 12. It indicates that the two imaging methods give similar cell analysis results. For instance, endothelial cell density shows similar numbers – 2902 cells/mm$^2$ for FD-FF-OCT and 2732 cells/mm$^2$ for a specular microscope. This demonstrates the utility of FD-FF-OCT for *in vivo* assessment of corneal endothelial cells. Since we had an ability to correct defocus in FD-FF-OCT images off-line, Visualization 4 shows segmentation performed for different defocus correction values. It demonstrates that we can numerically adjust images for the maximum contrast and the most optimal segmentation.

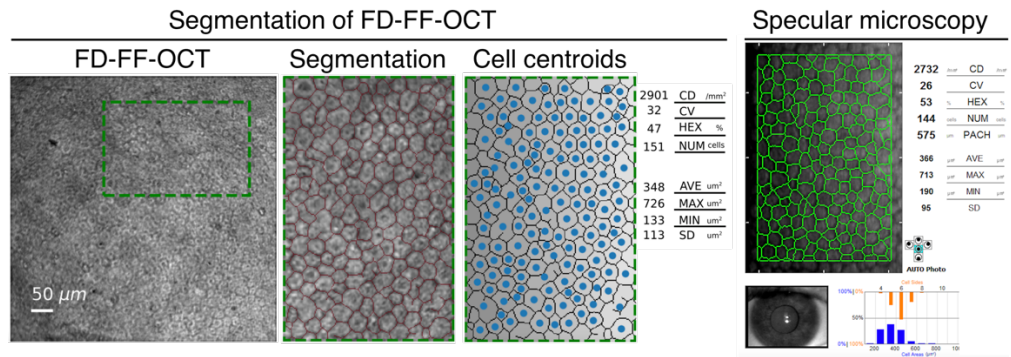

Fig. 12. Comparison of segmented FD-FF-OCT and specular microscopy images. CD – endothelial cell density; CV – coefficient of value (polymegathism); HEX – hexagonal cell ratio; NUM – total cell number; AVE – average cell area; MAX – the maximal cell area; MIN – the minimal cell area; SD – standard deviation of the mean cell area. Visualization 4 shows dynamic refocusing of endothelial cells and their corresponding segmentation for each refocus position.

**5. Discussion**

Recent developments in high-speed and high-FWC cameras enabled retinal imaging with TD-FF-OCT [24, 25] and FD-FF-OCT [28-30, 38, 41], but cornea was successfully imaged only with TD-FF-OCT [23]. Here, we showed that FD-FF-OCT could also be used to image various layers of the cornea in extremely short times when the spatial coherence of the swept laser source is destroyed to mitigate the laser safety constraints. FD-FF-OCT can currently achieve A-scan rates of 40 MHz and a voxel rates of around 10 GHz [28-30], making it the fastest OCT technique. Therefore, the FD-FF-OCT system presented here compares favorably to the other two state-of-the-art high-resolution and fast OCT systems for *in vivo* human corneal imaging – TD-FF-OCT [23] and FD(SD)-OCT [13]. Reported images of endothelial cells with TD-FF-OCT [23] were limited to approximately 100 μm × 200 μm, despite the overall FOV being 1.26 mm × 1.26 mm, which was due to the *en face* imaging of curved surface. We show in Figs 10 and 11 at least 10 times larger area of endothelial cells, despite four times smaller overall FOV (of 0.61 mm × 0.61 mm). Imaging with effectively larger FOV was possible because FD-FF-OCT acquired volumetric data that could be straightened up per individual layer. It is nearly impossible to acquire 3D volumes of the cornea with TD-FF-OCT because of eye movement and speed limitations – the axial images are blurred-out in *in vivo* imaging situations. Therefore, only *en face* images have been recorded so far and no B-scans (axial images) reported with TD-FF-OCT. We also show in Fig. 8 that FD-FF-OCT can image epithelium cells, which could not be well resolved with TD-FF-OCT [23], despite our system having slightly lower lateral resolution (2.4 µm versus 1.7 µm in Ref. [23]). The lower quality images of epithelium can be explained by motion artifacts in TD-FF-OCT that appear between consecutive frames. Also, strong specular reflections lowers image contrast but those could be suppressed by implementing dark-field detection [47]. We were able to remove the axial and lateral motion artifacts in FD-FF-OCT computationally. Specifically, phase information was used to suppress axial motion and registration of volumetric (3D) data eliminated lateral shifts. Some superficial epithelial cells can be seen when TD-FF-OCT is combined with FD(SD)-OCT [48] for the real time optical pathlength matching, reducing axial misalignment. FD-FF-OCT can correct for optical aberrations computationally and recover OCT signal. Even though TD-FF-OCT is less sensitive to optical aberrations [24, 49], it does, however, decrease OCT signal that cannot be recovered, especially in retinal imaging [24]. FD-FF-OCT is also capable of correcting for chromatic dispersion numerically, making the physical dispersion compensation less necessary, whereas TD-FF-OCT requires accurately matching dispersion physically between the reference and sample arms, which also changes with the imaging depth and therefore has to be dynamically adjusted [50]. Although TD-FF-OCT is an inherently less expensive technique, however, for *in vivo* imaging it needs a better axial tracking system of cornea, like one implemented in Ref. [48] with the SD-OCT add-on, making it, however, more expensive and sophisticated.

FD-OCT corneal imaging system reported in Ref. [13] had slightly larger FOV (of 0.75 mm × 0.75 mm) and more than twice better axial resolution (5.6 µm versus 2.1 µm in Ref. [13]) than our FD-FF-OCT system. However, FD-OCT system was an order of magnitude slower – recording the whole volume in 2.8 s – despite using one of the fastest spectrometers commercially available (of 150 kHz). Furthermore, it was not possible to perform DAC on the volumes acquired with the FD-OCT system, whereas in FD-FF-OCT it was feasible because of the laterally stable phase even when it was scrambled by DM before the interferometer [42].

Cornea is a good target for FD-FF-OCT due to its low scattering and high transparency that makes the confocal pinhole redundant. Crosstalk, appearing due to scattering, should also be less of an issue compared to skin [39] or retina [41] *in vivo* imaging with FD-FF-OCT. However, we could not determine how much of the crosstalk was suppressed by using DM in this work due to the laser safety constraints that made it impossible to acquire images with spatially coherent illumination at full power. To solve that, images with both coherent and incoherent illumination could be recorded simultaneously by shifting DM laterally, as in Ref. [42], so that part of the cornea would be illuminated with the modulated segment of DM and part with the unmodulated one, retaining retina exposure below the MPE. Nevertheless, as can be seen by comparing images in Fig. 7(a) and Fig. 7(b), the suppression of the laser coherence led to the reduction of the coherent artifacts in FF-OCT images. Thus, the reduction of coherence might generally improve the image quality and not only increase in the spot size on the retina, as was illustrated in Fig. 2.

The reasoning behind the current optical design was that, on the one hand, it was established that at least a 100 Hz sweep rate was required for *in vivo* imaging [36], which converts to the camera frame rate of at least 60 kHz when 512 images per sweep are recorded. On the other hand, FOV should be at least 0.5 mm × 0.5 mm to capture enough corneal details. Since the camera can acquire images with 512 × 512 pixels at 60 kHz, the FOV of 0.5 mm × 0.5 mm on the cornea can be therefore sampled approximately every 1 µm. Considering the Nyquist criterion, the resulting spatial resolution is then 2 µm, which can be traded-off against the FOV. The largest possible FOV for this particular objective was 2.65 mm × 2.65 mm, which follows from $FOV = \frac{FN}{M}$, where $M = \times 10$ and $FN = 26.5\ mm$, as specified by the manufacturer. Such large FOV could be recorded only if the camera had $4.3^2$ times more pixels than the current one or if the magnification on the camera was reduced 4.3 times at the expense of lowering the lateral resolution by the same factor of 4.3 (to ~ 10 μm). However, despite recent progress in camera development, cameras with such specifications do not yet exist, and lowering the lateral resolution was not desirable here.

Improvements in swept laser technology, such as in spectral bandwidth and power, would also be beneficial. Namely, a larger sweeping bandwidth would result in better axial resolution and would enable seeing the Descemet's and Bowman's membranes. Furthermore, higher power would allow to lower the reference reflectivity, which currently is ~10 %, and therefore, detect more photons from the cornea that would, in turn, increase SNR if the camera is operated close to saturation. Certain designs could also help detecting more photons through a more efficient interferometer [51] or beamsplitter [52] designs. Those configurations, if properly aligned, can also function in the dark-field mode rejecting specular reflections from the cornea. However, since cornea acts as the convex lens with ~4 mm focal length, it refocuses the specular reflections behind the pupil plane [52], where normally a block suppressing reflections is placed. Thus, cornea shape has to be considered when designing such a system.

Finally, extension of DOF would help to acquire the entire thickness of cornea. This can be attempted either by further manipulating phase computationally [53] or, for example, through additional optical elements that modifies phase in the pupil plane.

## 6. Summary

We have demonstrated high volumetric speed and spatial resolution human corneal imaging *in vivo* with Fourier-domain FF-OCT. Specifically, data volumes with FOV of 615 μm × 615 μm could be recorded in 8.6 milliseconds with the spatial resolution of 2.4 μm (laterally) and 5.6 μm (axially). Volume averaging was used to produce high contrast images of various corneal layers. Spatial coherence suppression with a fast-deformable membrane prevented the laser focusing on the retina to a spot with the intensity below the safety limit. We thus were able to acquire a significant portion of cornea's thickness with various corneal layers, such as epithelium, stroma and endothelium. Subbasal and middle stroma nerves were also seen. The results show that FD-FF-OCT has potential clinical value for non-invasive diagnostics of corneal diseases and its treatment monitoring.

## Funding

National Science Center (NCN, 2016/22/A/ST2/00313); European Union's Horizon 2020: research and innovation programme (666295); Polish Ministry of Science and Higher Education (2016-2019 int. co-financed project).

## Disclosure

The authors declare no conflicts of interest.